\documentclass[pre,twocolumn,aps,showpacs,showkeys]{revtex4}
\usepackage{graphicx}

\begin{document}
\title{Magnetoinductive breathers in magnetic metamaterials
}
\author{M. Eleftheriou$\ ^{1,2}$, N. Lazarides$\ ^{1,3}$, and G. P. 
Tsironis$\ ^{1}$
}
\affiliation{
$\ ^{1}$Department of Physics, University of Crete, 
and Institute of Electronic Structure and Laser,
Foundation for Research and Technology-Hellas,
P. O. Box 2208, 71003 Heraklion, Greece \\
$\ ^{2}$Department of Music Technology and Acoustics, Technological 
Educational Institute of Crete, E. Daskalaki, Perivolia, 74100 Rethymno, 
Crete, Greece \\
$\ ^{3}$Department of Electrical Engineering, Technological Educational 
Institute of Crete, P. O. Box 140, Stavromenos, 71500, Heraklion, Crete, 
Greece
}
\date{\today}

\begin{abstract}
The existence and stability of discrete breathers (DBs) in one-dimensional and 
two-dimensional magnetic metamaterials (MMs), which consist of periodic arrangements
(arrays) of split-ring resonators (SRRs), is investigated numerically. 
We consider different configurations of the SRR arrays,
which are related to the relative orientation of the SRRs in the MM,
both in one and two spatial dimensions. In the latter case we also consider
anisotropic MMs. Using standard numerical methods
we construct several types of linearly stable breather excitations both in 
Hamiltonian
and dissipative MMs (dissipative breathers).
The study of stability in both cases is performed using standard Floquet analysis.
In both cases we found that the increase of dimensionality from one
to two spatial dimensions does not destroy the DBs, which may also exist 
in the case of moderate anisotropy (in two dimensions).
In dissipative MMs, the dynamics is governed by a power balance between 
the mainly Ohmic dissipation and driving by an alternating magnetic field.
In that case it is demonstrated that DB excitation locally alters
the magnetic response of MMs from paramagnetic to diamagnetic. 
Moreover, when the frequency of the applied field approaches the SRR
resonance frequency, the magnetic response of the MM in the region of 
the DB excitation may even become negative (extreme diamagnetic).
\end{abstract}

\maketitle

\section{Introduction
}
Discrete breathers (DBs), also known as intrinsic localized modes (ILMs),
are genuine
nonlinear excitations that oscillate for long times in a localized region
of space and  may be produced generically in discrete lattices of
weakly coupled nonlinear elements
(see \cite{Flach} for a general review).
Since their introduction \cite{Sievers}, a large volume of analytical and
numerical studies have explored the existence and the properties of DBs
in a variety of discrete nonlinear systems.
Rigorous mathematical proofs of existence of DBs in both
Hamiltonian (i.e., energy conserved) and dissipative lattices of weakly
coupled nonlinear oscillators have been given \cite{Mackay,Aubry},
and numerical algorithms for their accurate construction
have been designed \cite{Marin,Marin1,Zueco}.
DBs may appear spontaneously in a lattice
as result of fluctuations \cite{Peyrard,TA,Rasmussen},
disorder \cite{Rasmussen1},
or by purely deterministic mechanisms \cite{Hennig}.
They have been observed experimentally in several systems, including
solid state mixed-valence transition metal complexes \cite{Swanson},
quasi-one dimensional antiferromagnetic chains \cite{Schwarz},
arrays of Josephson junctions \cite{Trias},
micromechanical oscillators \cite{Sato}, optical waveguide systems \cite{Eisenberg},
and proteins \cite{Edler}.
Once generated, DBs modify system properties such as lattice thermodynamics
and introduce the possibility of nondispersive energy transport 
\cite{Tsironis,Kopidakis},
because of their potential for translatory motion (i.e., mobility)
along the lattice \cite{Flach1}. 
In numerical experiments DB mobility can be achieved 
by an appropriate perturbation \cite{Chen}.
Recently, it has been found experimental evidence for moving DBs in a
layered crystal insulator at $300 K$ \cite{Russell}
and in a small micromechanical cantilever array \cite{Sato1}.
From the perspective of applications to experimental situations
where dissipative effects are always present, dissipative DB excitations
(usually driven by an alternating power source) 
are more relevant than their Hamiltonian counterparts.
Dissipative DBs, which possess the character of an attractor for initial
conditions in the corresponding basin of attraction,
may appear whenever power balance, instead of energy conservation,
governs the nonlinear dynamics of the lattice.
Furthermore, the attractor character of dissipative DBs allows for the existence
of quasiperiodic and even chaotic DBs \cite{Martinez1,Martinez}.

Recently, dissipative DBs have been demonstrated numerically in discrete
and nonlinear magnetic metamaterials (MMs)
driven by an alternating electromagnetic (EM) field \cite{Lazarides}.
The MMs are artificially structured materials that exhibit
EM properties not available in naturally occurring materials. 
The response of any material (either natural or artificial)
to applied EM fields is characterized by macroscopic parameters such
as the electric permittivity $\epsilon$ and the magnetic permeability $\mu$.
For example, there are only a few natural materials responding magnetically
at Terahertz (THz) and optical frequencies, and that response is usually
very weak.  However the MMs exhibit relatively
large magnetic response at those frequencies \cite{Yen,Podolskiy,Soukoulis},
which may be either positive 
or negative, resulting in positive or negative $\mu$, respectively.
The realization of MMs at such frequencies will certainly affect THz
optics and its applications, while it promises new device applications.
The most common element utilized for magnetic response from MMs is the 
split-ring resonator (SRR) which, in its simplest form, is a
metallic ring with one slit, made of a highly conducting metal.
Periodic SRR arrays in the nanoscale, a genuine realization of MMs,
are fabricated 
routinely using conventional microfabrication techniques.
These MMs, in which the SRRs are weakly coupled magnetically through 
their mutual inductances, support a new type of guided waves,
the magnetoinductive (MI) waves 
\cite{Shamonina,Freire,Sydoruk,Syms}.
The MI waves propagate within a band near the resonant frequency
of the SRRs, and exist as forward and backward waves depending 
on the orientation of the elements (e.g., the SRRs) of the MM.
In the linear regime of MI wave propagation in a MM, the magnetic 
permeability $\mu$ does not depend on the intensity of the EM field. 
However, the MMs may become nonlinear,
either by embedding the SRRs in a Kerr-type medium \cite{Zharov,OBrien},
or by inserting certain nonlinear elements (e.g., diodes) in each 
SRR \cite{Lapine,Shadrivov2,Shadrivov1}.
Then, the combined effects of nonlinearity and discreteness
(inherent in SRR-based MMs), leads in the generation of nonlinear excitations
in the form of DBs \cite{Lazarides},
as well as magnetic domain walls \cite{Shadrivov},
and magnetoinductive envelope solitons \cite{Kourakis}.
The latter may be both bright or dark, and they result from the 
modulational instability of the MI wares.
While the nonlinearity by itself offers tunability, 
stationary (i.e., not mobile, pinned) DBs act as stable impurity modes
that are dynamically generated and may alter propagation
and emission properties of a system. 
Moreover, stationary dissipative DBs can change locally the magnetic 
response of a nonlinear MM \cite{Lazarides}.
We should also note that regular arrays of rf SQUIDs offer an alternative 
for the construction of 
nonlinear MMs due to the nonlinearity of the Josephson 
junction \cite{Lazarides1}. 

In the present work we investigate numerically the existence and stability
of both dissipative
and energy conserved DBs in discrete, periodic arrays of nonlinear SRRs. 
We consider different SRR array geometries 
(i.e., different orientations of the SRRs in the MM)
in one and two spatial dimensions. 
In two dimensions we also consider several cases of possible anisotropy.
In the next section we describe the discrete MM model, while in section III we 
construct several types of DBs for one-dimensional (1D) arrays.
Here we also calculate the magnetic response, which may be locally 
altered by the presence of a DB.
In section IV we construct several types of DBs for two-dimensional (2D)
arrays,
and we finish in section V with the conclusions.

\section{Magnetic metamaterial model 
}
We consider 1D and 2D discrete, periodic
arrays of identical nonlinear SRRs, which consist the simplest realization
of a MM in one and two dimensions \cite{Katsarakis}, respectively.
In one dimension, the SRRs form a linear array with their centers separated
by distance $d$.
In two dimensions, the SRRs are assumed to be arranged in a regular
rectangular
lattice, with their centers separated by distance  $d_x$ and $d_y$  
in the $x-$ and $y-$directions, respectively 
(i.e., lattice constants $d_x$ and $d_y$, respectively).
There are two configurations of interest in each case,
shown schematically in Figs. 1 and 2 for the 1D and 2D arrays, respectively.
A 1D array can be constructed either in the planar configuration,
where all SRR loops are in the same plane with their centers lying 
on a straight line (Fig. 1(a)), or in the axial configuration, where 
the line connecting the centers of the SRR loops is perpendicular 
to the plane of the loops (Fig. 1(b)).
Similarly, a 2D array can be constructed either in the planar configuration,
where all SRR loops are in the same plane with their centers
located on an orthogonal lattice
(Fig. 2(a)), or in the planar-axial configuration, where all SRRs
have the planar configuration in one direction  
while they have the axial configuration in the other direction (Fig. 2(b)).
%%%%----------figure1--------
\begin{figure}[!ht]
\includegraphics[angle=0, width=1. \linewidth]{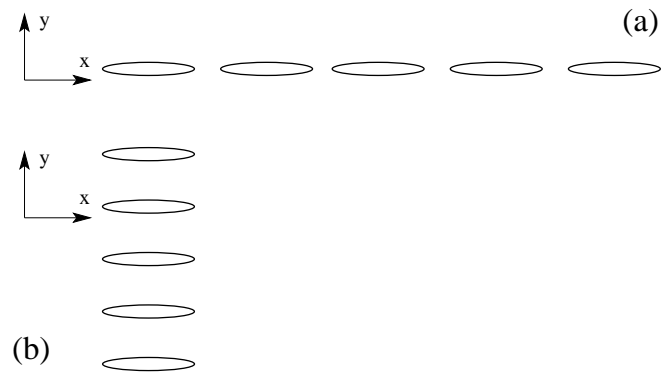}
\caption{
Schematic view of a one-dimensional array of split-ring resonators
in (a) the planar geometry; (b) the axial geometry.
In both geometries the split-ring resonator axes as well as the 
magnetic component of the applied field are directed along the 
$y-$axis.
The electric field component is transversal to the slits (not shown 
in the figure).
}
\label{Fig1}
\end{figure}

Within good approximation, each SRR is equivalent to a nonlinear
resistor-inductor-capacitor (RLC) circuit featuring a self-inductance $L$, 
Ohmic resistance $R$, and capacitance $C$. 
The units (i.e., the SRRs) become nonlinear due to a Kerr dielectric
filling the SRRs' slits, whose permittivity $\epsilon$ is of the form 
\begin{eqnarray}
\label{0}
   \epsilon (|{\bf E}|^2) = \epsilon_0 \left( \epsilon_\ell + \alpha 
      \frac{|{\bf E}|^2}{E_c^2} \right) ,
\end{eqnarray}
where ${\bf E}$ is the electric field,
$E_c$ is a characteristic (large) electric field, 
$\epsilon_\ell$ is the linear  permittivity,
$\epsilon_0$ is the permittivity of the vacuum, 
and $ \alpha=+1~~(-1)$
corresponding to self-focusing (self-defocusing) nonlinearity.
As a result, the SRRs acquire a field-dependent capacitance
$C ( |{\bf E}|^2 ) = \epsilon ( |{\bf E}_g|^2 )\, A / d_g$, 
where $A$ is the area of the cross-section of the SRR wire,
${\bf E}_g$ is the electric field induced along the SRR slit,
and $d_g$ is the size of the slit. 
The origin of ${\bf E}_g$ may be due to the magnetic 
and/or the electric components of the applied EM field,
depending on the relative orientation of that field with respect 
to the SRRs' plane and the slits.
In the following we assume that, for any array configuration and 
number of dimensions, the magnetic component of
the incident (applied) EM field is always perpendicular to the SRRs'
plane, and that the electric component of the incident EM field 
is transversal to the slit.
Then, only the magnetic component excites an electromotive force (emf)
in the SRRs, resulting in an oscillating current in each SRR
loop and the corresponding development of an oscillating voltage 
difference $U$
across the slits or, equivalently, of an oscillating electric 
field ${\bf E}_g$ in the slits.
%%%%----------figure2---------
\begin{figure}[!ht]
\includegraphics[angle=0, width=1. \linewidth]{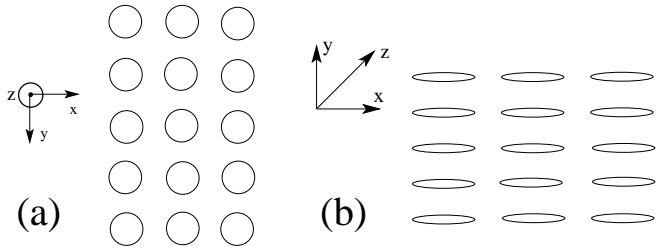}
\caption{
Schematic view of a two-dimensional array of split-ring resonators
in (a) the planar geometry; (b) the planar-axial geometry.
In both geometries the magnetic component of the applied field is
directed along the SRR axes,
while the electric field component is transversal to their slits 
(not shown in the figure). 
}
\label{Fig7}
\end{figure}
If $Q$ is the charge stored in the capacitor of an SRR then, 
from the general relation of a voltage-dependent capacitance,
$C(U)=dQ / dU$, and Eq. (\ref{0}), we get
\begin{eqnarray}
\label{1}
   Q = C_\ell 
     \left( 1 + \alpha \frac{U^2}{3 \epsilon_\ell \, U_{c}^2} 
     \right) U,
\end{eqnarray}
where $U =d_g E_g$,
$C_\ell = \epsilon_0 \epsilon_\ell (A / d_g)$ is the linear capacitance, 
and $U_{c} =d_g E_{c}$. 
Assume that the arrays are placed in a time-varying and spatially uniform
magnetic field of the form  
\begin{eqnarray}
\label{01}
  H = H_0 \, \cos(\omega t) ,
\end{eqnarray}  
where $H_0$ is the field amplitude, $\omega$ is the field frequency,
and $t$ is the time variable. The excited emf ${\cal E}$ , 
which is the same in all SRRs, is given by
\begin{eqnarray}
\label{02}
  {\cal E} = {\cal E}_0 \, \sin(\omega t), 
  \qquad {\cal E}_0 \equiv \mu_0 \, \omega \, S \, H_0 ,
\end{eqnarray}  
where $S$ is the area of each SRR loop, and $\mu_0$ the permittivity 
of the vacuum.
Each SRR in the field given by Eq. (\ref{01}) is a nonlinear 
oscillator which exhibits a resonant magnetic response (either positive
or negative) at a particular frequency which is very close to its linear
resonance frequency $\omega_\ell = 1/ \sqrt{L \, C_{\ell}}$
(for $R\simeq 0$).

All SRRs in an array are coupled together due to magnetic dipole-dipole 
interaction through their mutual inductances. However, we assume below
only nearest neighbor interactions, so that neighboring SRRs are coupled
through their mutual inductances $M_x$ and $M_y$. This is a very good
hypothesis in the planar configurations (i.e., both in 1D and 2D arrays),
even if the SRRs are very close. The validity of 
the nearest neighbor approximation for the other configurations
(i.e., the axial configuration in 1D arrays and the planar-axial 
configuration in 2D arrays) has been checked by taking into account 
the interaction of the SRRs with their four nearest neighbors,
using that the mutual 
inductance $M_{x,y}^{(s)}$ between an SRR and its $s-$th neighbor 
goes like $M_{x,y}^{(s)} \simeq M_{x,y} / s^3$. We found that for weak coupling,
as we consider here, the results are practically the same with those
obtained with the nearest neighbor approximation.
Thus, the electrical equivalent of an SRR array in an alternating magnetic field
is that of nonlinear RLC oscillators coupled with their nearest neighbors
through their mutual inductances, which are driven by identical alternating
voltage sources.
Therefore the equations describing the dynamics of the charge $Q_{n,m}$ 
and the current $I_{n,m}$ 
circulating in the $n,m-$th SRR may be derived simply from Kirchhoff's
voltage law for each SRR \cite{Lazarides,Shadrivov}
\begin{eqnarray}
\label{2}
  \frac{dQ_{n,m}}{dt} &=& I_{n,m}  \\  
\label{3}
    L \frac{dI_{n,m}}{dt} &+& R I_{n,m} + f (Q_{n,m})=
    \nonumber \\
   &-& M_{x} \left(\frac{dI_{n-1,m}}{dt}+\frac{dI_{n+1,m}}{dt} \right)
    \nonumber \\
   &-& M_{y} \left(\frac{dI_{n,m-1}}{dt}+\frac{dI_{n,m+1}}{dt} \right)
 	+ {\cal E},
\end{eqnarray}
where  $f(Q_{n,m})=U_{n,m}$ is given implicitly from Eq. (\ref{1}). 
Using the relations 
\begin{eqnarray}
\label{4}
  \omega_\ell^{-2} &=& L  C_\ell,~~\tau=t  \omega_\ell,
  ~~I_c = U_c  \omega_\ell  C_\ell,~~Q_c=C_\ell  U_c \\
  \label{5}
 {\cal E} &=& U_c \varepsilon,~~I_{n,m}=I_c i_{n,m},~~Q_{n,m} = Q_c q_{n,m} , 
\end{eqnarray} 
and Eq. (\ref{02}), Eqs. (\ref{2}) and  (\ref{3}) can be normalized to
\begin{eqnarray}
\label{6}
   \frac{d q_{n,m}}{d\tau} &=& {i_{n,m}} \\
\label{7}
  \frac{d i_{n,m}}{d\tau} &+&\gamma \, i_{n,m} + f (q_{n,m}) + 
   \lambda_{x} \left( \frac{d i_{n-1,m}}{d\tau} +\frac{d i_{n+1,m}}{d\tau} 
        \right)
	\nonumber \\
  &+&\lambda_{y} \left( \frac{d i_{n,m-1}}{d\tau} +\frac{d i_{n,m+1}}{d\tau} 
        \right)
    = \varepsilon_0 \, \sin(\Omega\tau) , 
\end{eqnarray}
where $\gamma=RC_{\ell}\omega_{\ell}$ is the loss coefficient, 
$\lambda_{x,y} =M_{x,y} / L$ are the the coupling parameters
in the $x-$ and $y-$direction, respectively, 
and $\varepsilon_0 = {\cal E}_0 / U_c$.
Note that the loss coefficient $\gamma$, which is usually very small 
($\gamma \ll 1$), may account both for Ohmic and radiative losses
\cite{Kourakis}.
The corresponding equations for 1D arrays result from Eqs. (\ref{6}) and 
(\ref{7}), i.e., by setting $\lambda_y=0$, dropping the subscript $m$,
and choosing the appropriate $\lambda_x = \lambda$.
Neglecting losses and without applied field,  Eqs. (\ref{6}) and (\ref{7}) 
can be derived from the Hamiltonian
\begin{eqnarray}
 \label{8}
  {\cal H} &=& \sum_{n,m} \left\{
      \frac{1}{2} \dot{q}_{n,m}^2  + V_{n,m} \right\}
      \nonumber \\
      &-&\sum_{n,m} \left\{
     \lambda_x \, \dot{q}_{n,m}\, \dot{q}_{n+1,m}
    +\lambda_y \, \dot{q}_{n,m}\, \dot{q}_{n,m+1}
      \right\}  ,
\end{eqnarray}
where the nonlinear on-site potential $V_{n,m}$ is given by
\begin{eqnarray}
 \label{9}
   V_{n,m} \equiv V( q_{n,m} ) =\int_0^{q_{n,m}} f(q_{n,m}') \, dq_{n,m}' ,
\end{eqnarray}   
and $\dot{q}_{n,m} \equiv d{q}_{n,m} / d\tau$.
Analytical solution of Eq. (\ref{1}) for $u_{n,m}=f(q_{n,m})$  
with the conditions of $u_{n,m}$ being real and 
$u_{n,m}(q_{n,m}=0)=0$, gives the approximate expression 
\begin{eqnarray}
\label{10}
   f(q_{n,m}) \simeq q_{n,m} -\frac{\alpha}{3\epsilon_\ell}q_{n,m}^{3} 
     + 3\left(\frac{\alpha}{3\epsilon_\ell}\right)^{2} q_{n,m}^{5} , 
\end{eqnarray}
which is valid for relatively low $q_n$ ($q_n < 1,~~n=1,2,...,N$)
Thus, the on-site potential is soft for $\alpha=+1$ and hard for $\alpha=-1$. 
In the 2D case the mutual inductances $M_{x}$ and $M_{y}$ may differ both 
in their sign, depending on the configuration, and their magnitude. 
Actually, even in the planar 2D configuration with $d_x=d_y$ a small 
anisotropy should be expected because we considered SRRs having only one slit.
This anisotropy can be effectively taken into account by considering 
slightly different
coupling parameters $\lambda_x$ and $\lambda_y$.
The coupling parameters  $\lambda_{x,y}$ as well as the loss coefficient 
$\gamma$ can be calculated numerically for this specific model within
high accuracy. However, for our purposes, it is sufficient to estimate 
these parameters for realistic (experimental) array parameters,
ignoring the nonlinearity of the SRRs and the effects due to the (weak) coupling.
The self-inductance of a circular SRR of radius $a$ (with circular cross-section
of diameter $h$) can be determined by the well-known 
expression $L = \mu_0 a [\ln(16a/h)-1.75]$. The value of the capacitance
then follows from the choice of the resonance frequency($\sim \omega_\ell$).
The resistance can be calculated from the 
given SRR dimensions and the appropriate value of the conductivity, 
taking into account the skin effect.
The expression for the mutual inductance between two loops can be calculated
by means of a simple approximation \cite{Lazarides}.
For example, for an array of circular silver-made SRRs with circular 
cross-section in the planar geometry, with geometrical and material parameters 
close to those in Ref. \cite{Katsarakis}
(for the equivalent squared SRR with squared cross-section), 
it has been estimated that $\lambda \approx 0.02$  
and $\gamma \approx 0.01$ \cite{Lazarides}.
For the same SRRs in the axial geometry, separated by distance $d$ as 
those in the planar geometry, the approximate expressions for the mutual
inductance and the coupling parameter are $M \simeq (\pi/2)\mu_0 a (a/d)^3$
and $\lambda \simeq (\pi/2)\mu_0 a (a/d)^3 /L$, respectively.
Those expressions are obtained by making the same approximations as
in Ref. \cite{Lazarides}, 
and knowing that the magnetic field of one of the SRRs with induced current $I$ 
at the center of the other SRR, which is directed along its axis, 
is given by $B = \mu_0 I a^2 / 2(a^2 + d^2)^{3/2}$.
Thus, in a first approximation, the coupling of two SRRs in the axial geometry
is stronger (by a factor of 2) than that of the same SRRs in the planar geometry.
%%%%----------figure3---------
\begin{figure}[!ht]
\includegraphics[angle=0, width=0.5 \linewidth]{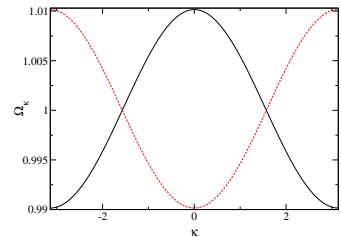}
\caption{
(Color online) 
 The frequency spectrum of linear MI waves $\Omega_{\kappa}$
as a function of the wavenumber $\kappa$, for  
$\lambda=-0.01$ (black-solid curve) and $\lambda=+0.01$ (red-dashed curve).
}
\label{Fig2}
\end{figure}

For the integration of Eqs. (\ref{6}) and (\ref{7}),
both in the Hamiltonian and the damped driven cases,
or the corresponding linearized ones, we use a standard fourth-order
Runge-Kutta algorithm with fixed time-stepping $\Delta t$
(typically  $\Delta t =0.01$).
Since the DBs studied here are highly localized,
the choice of boundary conditions to be imposed on Eqs. (\ref{6}) and (\ref{7})
is not especially important.
Thus, we have chosen periodic boundary conditions throughout the study. 
	
\section{Discrete breathers in one-dimensional SRR arrays
}
We consider the two different 1D configurations of SRR arrays shown in 
Fig. 1, with the same number of SRR oscillators $N=50$. 
Within the equivalent circuit model, the difference between
the two configurations resides in the sign and the magnitude of the 
coupling parameter $\lambda$.  
In the planar geometry (configuration) $\lambda$ is negative, 
since the mutual inductance $M$ between two neighboring SRRs 
is negative.  This is due to the fact that the magnetic field
generated by one SRR (due to the current induced in its loop)
crosses the neighboring SRR in the opposite direction.
For the axial geometry, the mutual inductance $M$, and hence 
$\lambda$ is positive.  Moreover, in a 1D array of SRRs in the axial 
geometry,  the value of $\lambda$ is much higher than that 
in a 1D array of SRRs in the planar geometry with the same SRR spacing $d$.
High coupling between SRRs in the axial geometry would possibly require to
take into account the interaction of an SRR with its far neighbors
(and not only with its nearest neighbors). 
To avoid such complications, and for the sake of comparison of the DBs
obtained in the two geometries, we assume that the SRR spacing is such that 
the magnitude of the coupling between neighboring SRRs is the same 
(or at least of the same order) in both geometries.

For Hamiltonian systems DBs may be constructed from the anti-continuous
limit \cite{Marin}, where all the SRRs are uncoupled
($\lambda \rightarrow 0$), obeying identical dynamical equations.
Fixing the amplitude of one of them (say the one located at $n=n_b$)
to a specific value $q_{b}$, with the corresponding current $i_{b}=0$,
we can determine, either analytically (if possible) or numerically,
the period of oscillation $T_b$.
An initial condition with $q_{n}=0$ for any $n\neq n_b$, $q_{n_b}=q_b$,
and $\dot{q}_{n}=i_n=0$ for any $n$, represents a trivial DB with period $T_b$.
Continuation of this solution for $\lambda\neq 0$ using the Newton's method
\cite{Marin},
results in numerically exact DBs up to a maximum value of the coupling 
parameter $\lambda= \lambda_{max}$.
For the existence of Hamiltonian DBs it is required that the DB frequency
$\omega_b = 2\pi /T_b$, as well as all its multiples, lie outside 
the linear dispersion band of MI waves.  The MI wave band is obtained 
by substituting a plane wave of the form $q_{n}=A \cos(\kappa n-\Omega\tau)$
into the linearized equations Eqs. (\ref{6}) and (\ref{7}) 
(with $\varepsilon=0, \gamma=0$)
\begin{eqnarray}
\label{11}
\Omega_\kappa = \frac{1}{\sqrt{ 1 +2\, \lambda \, \cos(\kappa) } } ,
\end{eqnarray}
where $\Omega =\omega/\omega_\ell$ is the normalized frequency, 
and $\kappa = k\, d$ is the  normalized wavenumber ($-\pi \leq \kappa \leq \pi$).
Typical dispersion curves are shown in Fig. 3 for both geometries.
For $|\lambda| \ll 1$ the band is very narrow
(bandwidth $\Delta\Omega \simeq 2|\lambda|$), so that the requirement 
for the existence of Hamiltonian DBs can be easily satisfied.
Clearly, the bandwidth increases with increasing $\lambda$.
The MI waves are forward in the axial configuration ($\lambda >0$)
with co-directional phase and group velocities, and backward in the 
planar configuration ($\lambda <0$), with phase and group velocities
in the opposite directions.
%%%%----------figure4---------
\begin{figure}[!t]
\includegraphics[angle=0, width=1. \linewidth]{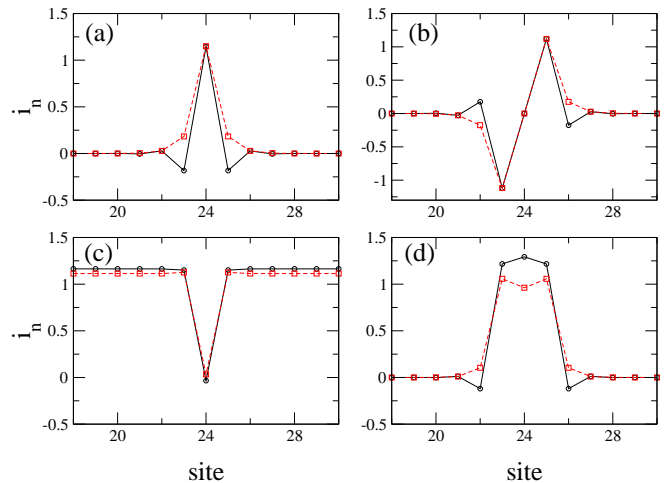}
\caption{(Color online) 
Several Hamiltonian discrete breather profiles (at maximum amplitude)
for $\alpha=+1$, $\Omega_{b}=0.938$, $\epsilon_\ell=2$, $N=50$,
constructed with Newton's method for both the planar and the axial
array configurations 
(shown as black circles and red squares, respectively.) 
(a) A single-site bright breather for $|\lambda|=0.02$;
(b) an antisymmetric bright breather for $|\lambda|=0.02$;
(c) a single-site dark breather for $|\lambda|=0.002$;
and (d) a bright multibreather for $|\lambda|=0.013$.
Only part of the simulated array is shown for clarity.
}
\end{figure}

Following the procedure described above,
with the appropriate choice of initial conditions (trivial DBs),
we have constructed several types 
of Hamiltonian, numerically exact DBs for 1D SRR arrays in both 
the planar and the axial geometries,
shown in Figs. 4 and 5 for self-focusing ($\alpha=+1$) and 
self-defocusing nonlinearities ($\alpha=-1$), respectively.
Those Hamiltonian DB profiles (shown at maximum amplitude), 
i.e., the normalized current $i_n$ as a function of array site $n$,
are characterized as 
single-site bright DBs (Figs. 4(a) and 5(a)), 
antisymmetric bright DBs (Figs. 4(b) and 5(b)), 
single-site dark DBs (Figs. 4(c) and 5(c)), 
and bright multibreather (Figs. 4(d) and 5(d)). 
The term "bright" ("dark") DBs is used when there are only one or a few
SRRs in which the current oscillates with large (small) amplitude,
whereas the rest of them oscillate with small (large) amplitude 
\cite{Alvarez}.
All these DBs are highly localized, occupying only a few lattice sites,
since they have been obtained for low values of the coupling 
parameter $\lambda$.
They are all symmetric, except those in Figs. 4(b) and 5(b) which are
anti-symmetric.
It is interesting to observe that, 
for self-focusing nonlinearity ($\alpha=+1$), 
the profile of the single-site bright DB (Fig. 4(a)) is staggered (unstaggered)
for SRR arrays in the planar (axial) geometry, 
while, for self-defocusing nonlinearity ($\alpha=-1$),
the profile of the single-site bright DB (Fig. 5(a)) is staggered (unstaggered)
for SRR arrays in the axial (planar) geometry. 
Recall that a staggered (unstaggered) DB is that whose profile is of the 
form $i_n = {\cal I}_n \, \exp[i \kappa_b (n-n_b)]$, ${\cal I}_n =|i_n|>0$,
with $\kappa_b=\pi$ ($\kappa_b=0$) \cite{Maluckov}.
The single-site DBs in Fig. 4(a) and 5(a), with (normalized) frequencies
$\Omega_b = \omega_b /\omega_\ell = 0.938$ and $1.056$, respectively,
may be continued for higher
values of coupling. They cease to exist when the MI wave band,
which expands with increasing $\lambda$ reaches the DB  frequency $\Omega_b$.
That will occur at $|\lambda| = |\lambda_{max}|=|1 -1/\Omega_b^2|/2$,
which gives $|\lambda_{max}| \simeq 0.068$ and $0.052$ for 
$\Omega_b = 0.938$ and $1.056$, respectively.

%%%%----------figure5---------
\begin{figure}[!ht]
\includegraphics[angle=0, width=1. \linewidth]{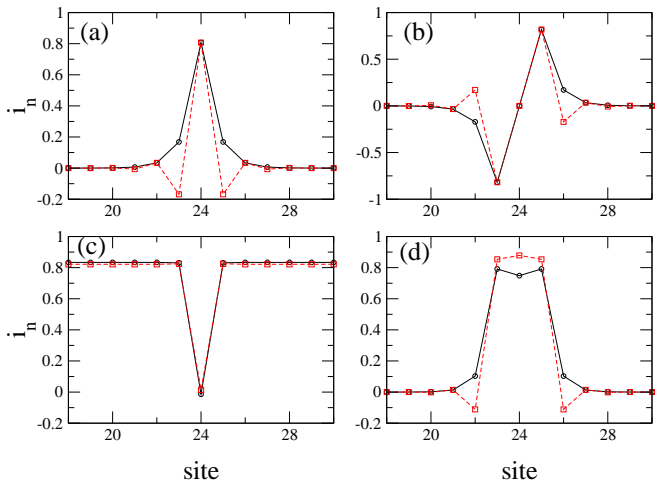}
\caption{(Color online) 
Several Hamiltonian discrete breather profiles (at maximum amplitude)
for $\alpha=-1$, $\Omega_{b}=1.056$, $\epsilon_\ell=2$, $N=50$,
constructed with Newton's method for both the planar and the axial
array configurations 
(shown as black circles and red squares, respectively.)
(a) A single-site bright breather for $|\lambda|=0.02$;
(b) an antisymmetric bright breather for $|\lambda|=0.02$;
(c) a single-site dark breather for $|\lambda|=0.001$;
and (d) a bright multibreather for $|\lambda|=0.013$.
Only part of the simulated array is shown for clarity.
}
\label{Fig4}
\end{figure}
The linear stability of Hamiltonian DBs is addressed
through the eigenvalues of the Floquet matrix (Floquet multipliers).
A DB is linearly stable when all its Floquet multipliers lie on a circle
of radius unity in the complex plane. 
The DBs shown in Figs. 4(a), 4(b) and 5(a), 5(b) are all linearly stable.
However, the dark DBs shown in  Figs. 4(c) and 5(c) as red squares, 
as well as the multibreathers shown in  Figs. 4(d) and 5(d) as red squares
and black circles, respectively, are linearly unstable.
Those DBs were found to be linearly stable only for very low couplings.
%%%%----------figure6---------
\begin{figure}[!ht]
\includegraphics[angle=0, width=0.8 \linewidth]{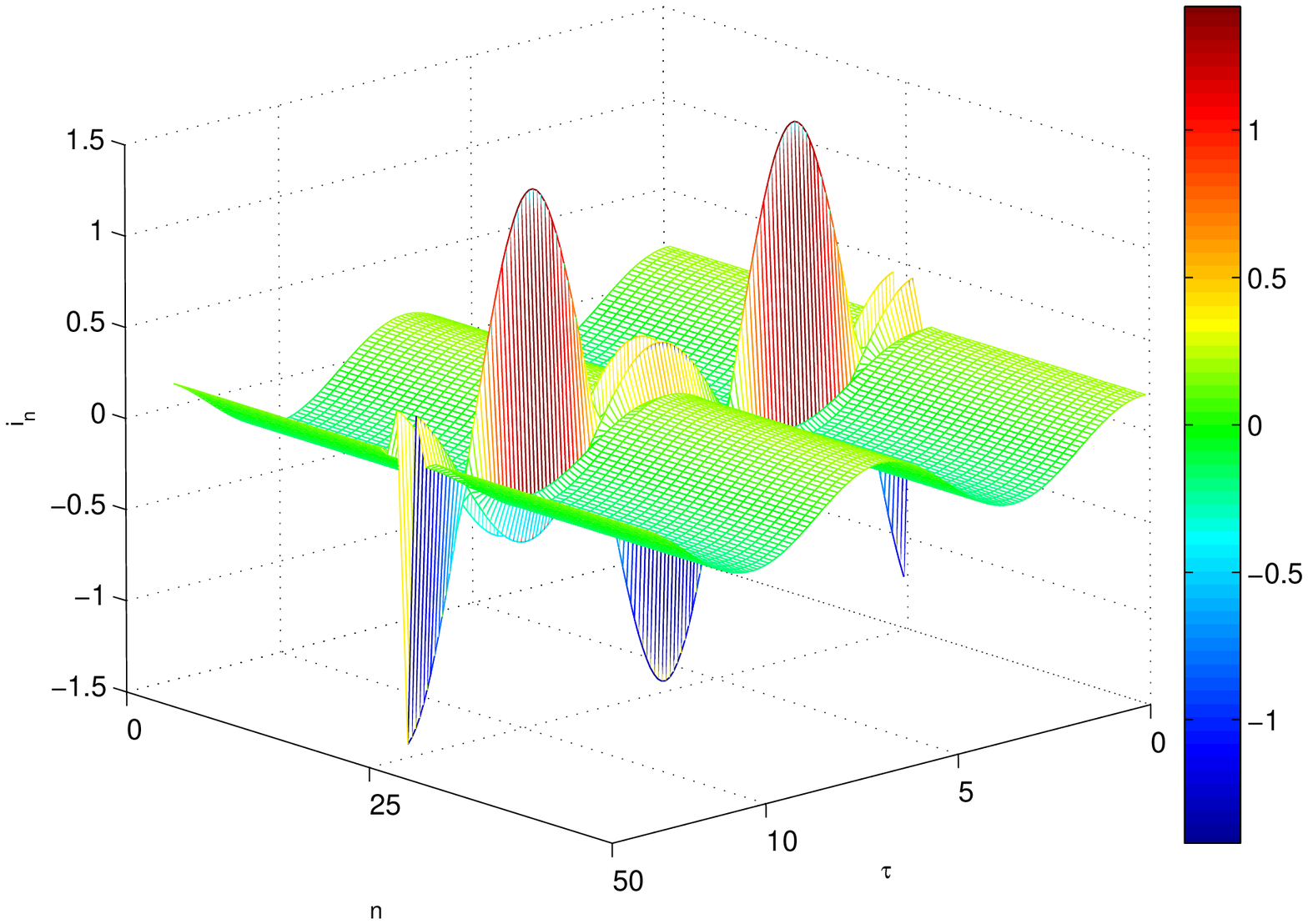}
\includegraphics[angle=0, width=0.8 \linewidth]{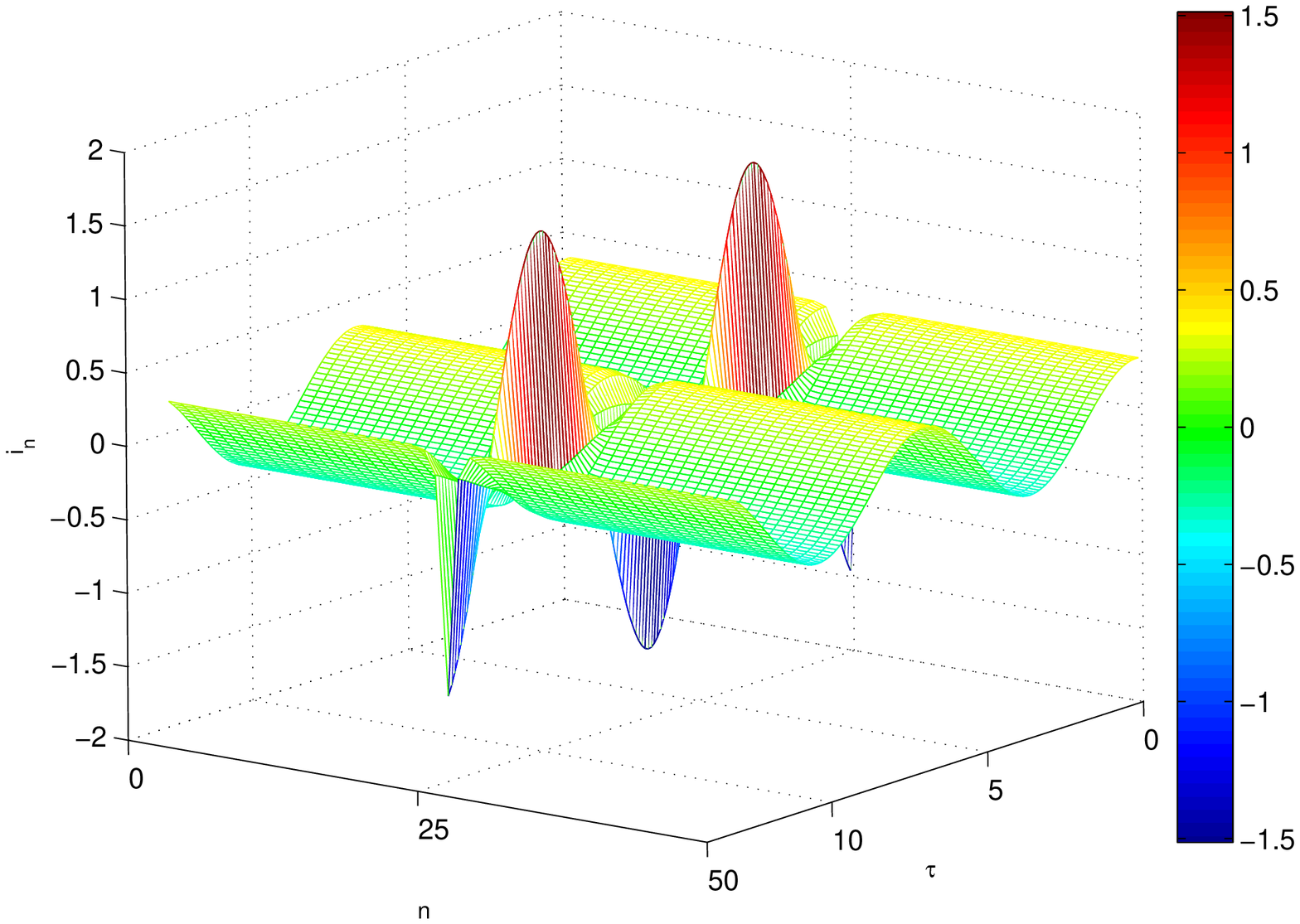}
\caption{(Color online) 
Time evolution of a single-site bright dissipative breather 
during approximately two periods,
for $\Omega_{b}=0.92$, $\varepsilon_0=0.04$, $\gamma=0.01$, 
$\epsilon_\ell=2$, $N=50$, and 
(top panel) planar SRR array geometry with $\lambda=-0.02$;
(bottom panel) axial  SRR array geometry with $\lambda=0.017$.
}
\end{figure}

Next, we construct DBs for SRR arrays which are subjected to losses
(damping)
and the external driving force $\varepsilon (\tau)$ (dissipative DBs).
In order to generate DBs in this case
we start by solving Eqs. (\ref{6}) and (\ref{7}) in the 
anti-continuous limit, where all SRRs are uncoupled.
We identify two coexisting (and stable) attractors of a single damped-driven
SRR oscillator
with focusing nonlinearity ($\alpha=+1$)
and $\Omega=0.92$, $\varepsilon_0=0.04$, $\gamma=0.01$,
which have different amplitudes $q_h \simeq 1.6086$  and $q_\ell \simeq 0.2866$
(high and low amplitude attractors, respectively).
Notice that the frequency $\Omega$ is below the SRR resonance frequency,
where a MM is expected to show only positive magnetic response
(i.e., response corresponding to positive $\mu$).
Subsequently, we fix the amplitude of one of the SRR oscillators
(say the one at $n=n_b$) to  $q_h$ and all the others to $q_\ell$
($i_n$ are all set to zero).
Using this configuration (trivial dissipative DB) as initial condition, 
we start integrating Eqs. (\ref{6}) and (\ref{7}), while increasing the 
coupling parameter $\lambda$ in small steps.
In this way, the initial condition can be continued for $\lambda \neq 0$
leading to dissipative DB formation \cite{Marin}.
The time evolution of typical dissipative (single-site) bright DBs is shown 
for SRR arrays 
in the planar and the axial geometries in Figs. 6
(top and bottom panels, respectively).
We see that both the DB and the background are oscillating with different
amplitudes (high and low, respectively). In this aspect, dissipative DBs differ
from their Hamiltonian single-site counterparts, where the background
is always zero.
We should note here that the DB frequency in this case has to be identical
with that of the driving field, so that $\Omega_b \equiv 2\pi/T_b = \Omega$.
However, the phase differences of the SRR oscillators in an array with respect
to the driving field are generally different, so that a DB may change 
locally the magnetic response of an array, as we shall see below.
With appropriate initial conditions we can also obtain  multi-breathers
where two or more sites oscillate with high (low) amplitude,
while the other ones with low (high) amplitude.

The linear stability of the dissipative DBs can be addressed 
(as in the Hamiltonian case) 
through the eigenvalues of the Floquet matrix (Floquet multipliers).
In the dissipative case, however, a DB is linearly stable when all
its Floquet multipliers lie on a circle of radius 
$R = \exp(-\gamma T_b /2)$ in the complex plane \cite{Marin1},
due to the presence of a dissipative term in the linearized 
equations of motion. Both the dissipative DBs shown in Fig. 6 are 
linearly stable. In order to check that result, we have also added   
small  perturbations to these DBs (of the order of $10^{-2}$ of the DB
amplitude) and let them evolve in time. 
We have followed the perturbed DBs over $10^{3}~T_b$ time units
observing that the DBs restore their unperturbed shape.
In general, dissipative DBs have been found to exist for couplings
$\lambda$ much less than those of their Hamiltonian counterparts,
$\lambda_{max}$. 
The DBs shown in Fig. 6 for dissipative and forced SRR arrays in the 
planar and the axial geometries, are found to exist up to 
$\lambda \simeq 0.024$ and $0.017$, respectively.
For defocusing nonlinearity 
($\alpha=-1$), it was impossible to identify two different amplitude attractors 
and, thus, we were not able to construct dissipative DBs in this case.
%%%%----------figure7---------
\begin{figure}[!ht]
\includegraphics[angle=0, width=0.85 \linewidth]{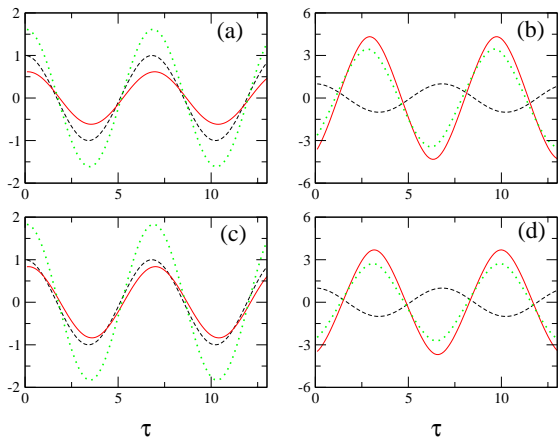}
\caption{(Color online) 
Time evolution of $\ell i_n(\tau)$ (red-solid curve), 
compared with $\cos(\Omega \tau)$ (black-dashed curve), and their sum 
(green-dotted curve), during two breather periods, for
(a) a 1D SRR array in the planar configuration at $n=15$ ($\lambda=-0.02$, $\ell=3$); 
(b) a 1D SRR array in the planar configuration at $n=n_b$ ($\lambda=-0.02$, $\ell=3$);
(c) a 1D SRR array in the axial configuration at $n=15$ ($\lambda=0.017$, $\ell=2.4$);
(d) a 1D SRR array in the axial configuration at $n=n_b$ ($\lambda=0.017$, $\ell=2.4$).
The other parameters as in Fig. 6.
}
\label{Fig6}
\end{figure}

It is  interesting to calculate the magnetization in a dissipative
SRR array. In the direction perpendicular to the SRR planes 
the general relation 
\begin{equation}
\label{12}
B=\mu_{0} ( H +{\cal M} )
\end{equation}
holds, with ${\cal M} = S \, I \, / d^3$ the magnetization 
(magnetic moment per unit cell volume) of the array. 
The earlier 
equation, with using Eq. (\ref{01}), can be written in normalized form as 
\begin{equation}
\label{13.1}
B/B_{0} = \cos(\Omega \tau)+ \ell \, i(\tau), 
\end{equation}
where $B_{0}=\varepsilon_0 U_c / S \, \omega$ and 
$\ell =\mu_0 \, S^2 \, \Omega / \varepsilon_0 \, d^3 \, L$. 
Eq. (\ref{13.1}) may be used locally at each cell, with the three 
terms $B/B_{0}$, $\cos(\Omega \tau)$, and $\ell i(\tau)$ representing
the instantaneous magnetic induction, applied magnetic field, and
magnetic response (local magnetization), respectively, in a specific cell.
Negative magnetic response appears, as it is apparent from Eq.(\ref{13.1}),
whenever the second term on the right-hand-side of Eq.(\ref{13.1})
larger in magnitude than the first one, and it has the opposite sign.
In Fig. 7 the time evolution is shown separately for each of the 
three terms of Eq.(\ref{13.1}), i.e., the quantity $\ell i_n(\tau)$ 
(red-solid curve), $\cos(\Omega \tau)$ (black-dashed curve), and their
sum $B/B_0$ (green-dotted curve), for two different sites
in both the planar and the axial geometries.
Specifically, Figs. 7(a) and 7(b) are for the planar geometry, for 
$n=15$ (a site on the background, quite far away from the central DB site)
and $n=n_b$ (DB central site), respectively, while
Figs. 7(c) and 7(d) are for the axial geometry,
for $n=15$ and $n=n_b$, respectively.  
We observe that, in both geometries,
the SRR with low amplitude current oscillation (reduced nonlinearity) shows 
positive (paramagnetic) response 
(Figs. 7(a) and 7(c) for the planar and the axial geometry, respectively),
while the SRR with high  amplitude current oscillation (enhanced nonlinearity) 
shows extreme diamagnetic (negative) response
(Figs. 7(b) and 7(d) for the planar and the axial geometry, respectively).
Thus, in the breather or multibreather location the lattice has a negative 
magnetic response even though it is driven below resonance.
That result can be extended to uniform solutions, where $I_n=I$ for all $n$.
Without nonlinearity such a solution always provides positive response
below the resonance frequency ($\sim \omega_\ell$).
However, the nonlinearity makes it possible to have two different coexisting
and stable states (bistability) with low and high amplitude current oscillation, 
which are in phase and in anti-phase, respectively, with the applied magnetic field.
Thus, it is possible to obtain from an SRR array uniform negative response below 
resonance,
by exciting all SRRs in the array to the high current amplitude state.
These remarks are also valid for 2D SRR arrays discussed in the next section.

\section{Discrete breathers in two-dimensional SRR arrays
}
Although most of the methodology and techniques for DB construction 
has been developed for the 1D case, there have been some studies of
higher-dimensional systems. 
Importantly, a rigorous proof of the existence of DBs in higher-dimensional
nonlinear lattices was given in Ref. \cite{Mackay}.
Numerical studies of DBs have been published for several simple 
nonlinear lattices as for example 2D Fermi-Pasta-Ulam chains 
\cite{Flach2,Marin2,Sankar,Butt},
Josephson Junction arrays \cite{Mazo}, Klein-Gordon chains \cite{Burlakov},
discrete nonlinear Schr\"odinger systems \cite{Kevrekidis,Gomez},
and also for a Morse lattice \cite{Ikeda}.
Since most of the present MMs are fabricated in 2D technology,
it is necessary to extend the study of magneto-inductive DBs in two 
dimensions. We see below that both the Hamiltonian and the dissipative 
DBs are not destroyed by increasing the dimensionality.

The frequency spectrum of linear MI waves in the 2D system can be obtained
as in the 1D case, by  substituting a plane wave of the form
$q_{n,m}=A \cos(\kappa_x n + \kappa_y m -\Omega\tau)$ into the 
linearized equations Eqs. (\ref{6}) and (\ref{7}) in the absence of losses 
and applied field ($\varepsilon_0 = 0, \gamma=0$)
\begin{eqnarray}
\label{13}
  \Omega_{\bf \kappa}  = 
  {[ 1 +2\, \lambda_{x} \, \cos(\kappa_x) 
       +2\, \lambda_{y} \, \cos(\kappa_y) ]}^{-1/2} ,
\end{eqnarray}
where $\kappa_x$ and $\kappa_y$ are the (normalized) components of the
wavevector in the $x-$ and $y-$direction, respectively
($-\pi \leq \kappa_i \leq \pi$, $i=x,y$).
In this case, the group velocity is not, in general, in a direction opposite 
to the phase velocity \cite{Syms}.
Notice that the bandwidth $\Delta \Omega$ depends both on the magnitude
of the coupling parameters $\lambda_x$ and $\lambda_y$ and their sign.
Consider, for example, the case $\lambda_x=\lambda_y =\lambda$ 
(isotropic lattice in the planar geometry). 
Then $\Delta \Omega \simeq 4|\lambda|$ for $|\lambda| \ll 1$,
which is larger than that of the corresponding 1D system by a factor of two.
Thus, in this case, the 1D Hamiltonian DBs with frequencies very close
to the 1D band may not survive in the 2D case. 
Typical dispersion curves (i.e., contours of the frequency as a function
of $\kappa_x$ and $\kappa_y$) of linear MI waves for isotropic 2D SRR arrays 
in the planar geometry, anisotropic 2D SRR arrays in the planar geometry,
and anisotropic 2D SRR arrays in the planar-axial geometry,  are shown 
in the left, middle, and right panels of Fig. 8, respectively.
%%%%----------figure8---------
\begin{figure}[!ht]
\includegraphics[angle=0, width=1. \linewidth]{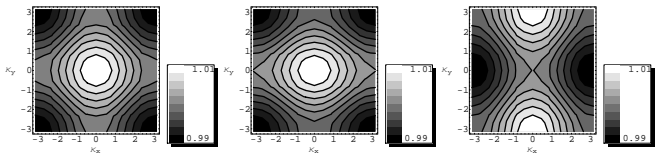}
\caption{
Dispersion curves of the frequency spectra of linear magnetoinductive waves
(along with their corresponding density plots) on the $\kappa_x - \kappa_y$
plane, for $\lambda_x=\lambda_y=-0.01$ (left panel);
$\lambda_{x}=-0.01, \lambda_{y}=-0.013$ (middle panel);
and $\lambda_{x}=-0.013, \lambda_{y}=0.01$ (right panel).
In the density plots, darker color indicates higher $\Omega_\kappa$.  
}
\end{figure}

For the construction of DBs in 2D arrays, we used the same methods as for the 
corresponding 1D arrays.  We again consider two different geometries
of the SRR arrays,
the planar geometry (Fig. 2(a)) and the planar-axial geometry (Fig. 2(b)).
In the former geometry, the coupling parameters $\lambda_x, \lambda_y$
are both negative. However, they may differ in magnitude, i.e.,
$|\lambda_x| \neq |\lambda_y|$ as a result of unequal lattice constants
$d_x$ and $d_y$ or resulting from the different orientation of the SRRs.
In the latter geometry, the coupling parameters have opposite signs,
i.e., $\lambda_x <0$ and $\lambda_y> 0$, which can be regarded as 
a case of generalized anisotropy. Their magnitude may however be different,
depending again on the lattice constants $d_x$ and $d_y$.
In the following we are mainly concerned with single-site
bright DBs for both geometries, nonlinearities 
(self-focusing and self-defocusing),
and several DB frequencies and pairs of coupling parameters 
$\lambda_x, \lambda_y$.
However, one may obtain many different types of DBs by just choosing the
appropriate initial condition (trivial DB). 
The 2D array size used in the calculations is typically $N\times N=15\times 15$.
%%%%----------figure9---------
\begin{figure}[!ht]
\includegraphics[angle=0, width=1. \linewidth]{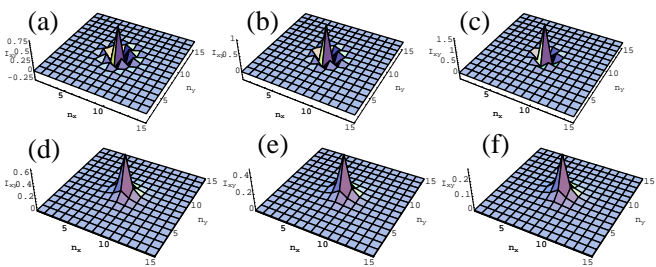}
\caption{(Color online) 
Snapshots of two-dimensional Hamiltonian single-site bright discrete breathers
(taken at maximum amplitude), for isotropic split-ring resonator arrays
in the planar geometry ($\lambda_{x}=\lambda_{y}=\lambda$) with
(a) $\alpha=+1$, $\Omega_b=0.952$, $\lambda=-0.020$;
(b) $\alpha=+1$, $\Omega_b=0.938$, $\lambda=-0.024$;
(c) $\alpha=+1$, $\Omega_b=0.881$, $\lambda=-0.040$;
(d) $\alpha=-1$, $\Omega_b=1.082$, $\lambda=-0.025$;
(e) $\alpha=-1$, $\Omega_b=1.036$, $\lambda=-0.012$; and
(f) $\alpha=-1$, $\Omega_b=1.011$, $\lambda=-0.004$.
}
\end{figure}

Typical Hamiltonian DB profiles in 2D isotropic ($\lambda_x=\lambda_y$)
SRR arrays 
in the planar geometry are shown in Fig. 9 for both self-focusing 
(Figs. 9(a) - 9(c)) and self-defocusing (Figs. 9(d) - 9(f)) nonlinearities.
The DB frequencies and coupling parameters are given, for each DB,
in the caption of Fig. 9.
Notice that for self-focusing nonlinearity ($\alpha=+1$), the DBs are 
staggered in both the $x-$ and the $y-$directions, while for
self-defocusing  nonlinearity ($\alpha=-1$), they are unstaggered
in both the $x-$ and the $y-$directions.
We have constructed exact DBs also in the case of an array in the planar 
geometry with moderate anisotropy, i.e., $\lambda_x \neq \lambda_y$,
for both nonlinearities ($\alpha=\pm 1$). Typical DB profiles 
with the coupling parameters differing by approximately $10\%$
are shown in Figs. 10(a) and 10(b) (upper panels) for $\alpha=+1$
and $\alpha=-1$, respectively. The anisotropy does not change the 
staggered/unstaggered character of these DBs, which remain staggered (unstaggered)
in both directions for $\alpha=+1$  ($\alpha=-1$).
The lower panels of Figs. 10(a) and 10(b) show the density plots of 
the corresponding
DB profiles shown in the upper panels.
In those density plots, where a darker color indicates a higher current amplitude,
a DB which is staggered in both directions appears as a checkerboard rotated 
by 45 degrees
with respect to the array.
%%%%----------figure10---------
\begin{figure}[!ht]
\includegraphics[angle=0, width=0.8 \linewidth]{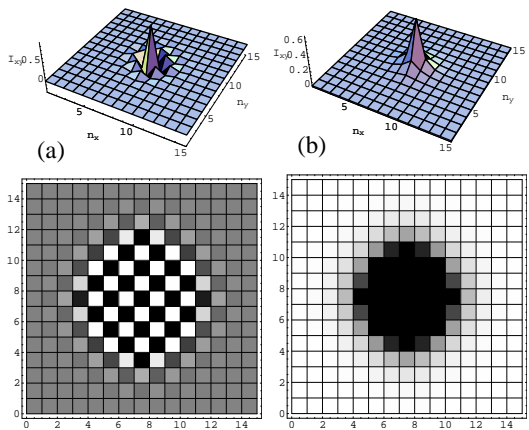}
\caption{(Color online) 
Snapshots of two-dimensional Hamiltonian single-site bright discrete breathers
(taken at maximum amplitude), for anisotropic split-ring resonator arrays
in the planar geometry with
(a) $\alpha=+1$, $\Omega_b=0.938$, $\lambda_x=-0.024$; $\lambda_y=-0.027$;
and
(b) $\alpha=-1$, $\Omega_b=1.082$, $\lambda_x=-0.025$, $\lambda_y=-0.028$.
In both (a) and (b), the lower panel show the density plot of the 
discrete breather profile presented in the corresponding upper panel.
In the density plots, darker color indicates higher breather amplitude.
}
\label{Fig10}
\end{figure}

Typical Hamiltonian DB profiles in 2D anisotropic ($|\lambda_x| \neq |\lambda_y|$) 
SRR arrays in the planar-axial geometry are shown in Figs. 11(a) and
11(b) (upper panels) for $\alpha=+1$ and for $\alpha=-1$, respectively,
along with their corresponding density plots (lower panels).
The coupling parameters differ in magnitude by approximately $10\%$.
In this case we observe that for $\alpha=+1$ ($\alpha=-1$) the DB is
staggered (unstaggered) along the $x-$direction ($y-$direction),
while it is unstaggered (staggered) along the $y-$direction ($x-$direction).
Thus, the change in either the sign of the nonlinearity $\alpha$ 
or the sign of the coupling parameter $\lambda_y$, leads to a change in the
staggered/unstaggered  character of a DB in the $y-$direction. 
Specifically,  
by changing $\alpha$ from +1 to -1 (for $\lambda_y >0$), a DB unstaggered 
in the $y-$direction becomes staggered in that direction,
while by changing $\alpha$ from -1 to +1 (for $\lambda_y >0$),
a DB staggered in the $y-$direction becomes unstaggered in that direction.
Also, by changing $\lambda_y$ from positive to negative (for  $\alpha=+1$),
a DB unstaggered in the $y-$direction becomes staggered in that direction,
while  by changing $\lambda_y$ from negative to positive (for  $\alpha=+1$),
a DB staggered in the $y-$direction becomes unstaggered in that direction.
The linear stability of all the Hamiltonian DBs presented in this section is checked
through Floquet analysis (finding the eigenvalues of the Floquet matrix),
and they were found to be linearly stable.
%%%%----------figure11---------
\begin{figure}[!ht]
\includegraphics[angle=0, width=0.8 \linewidth]{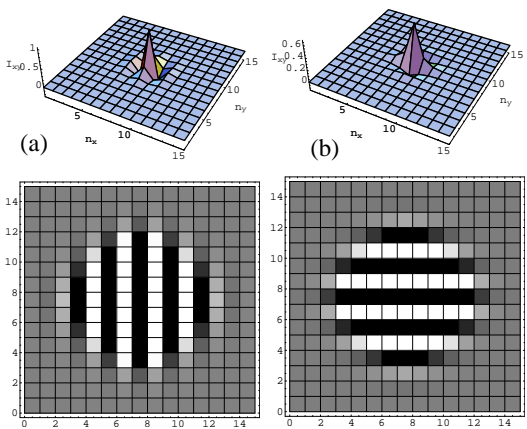}
\caption{(Color online) 
Snapshots of two-dimensional Hamiltonian single-site bright discrete breathers
(taken at maximum amplitude), for anisotropic split-ring resonator arrays
in the planar-axial geometry with
(a) $\alpha=+1$, $\Omega_b=0.938$, $\lambda_x=-0.024$; $\lambda_y=0.021$;
and
(b) $\alpha=-1$, $\Omega_b=1.082$, $\lambda_x=-0.025$, $\lambda_y=0.022$.
In both (a) and (b), the lower panel show the density plot of the 
discrete breather profile presented in the corresponding upper panel.
In the density plots, darker color indicates higher breather amplitude.
}
\label{Fig11}
\end{figure}

Typical examples of dissipative DB profiles in 2D isotropic ($\lambda_x=\lambda_y$)
SRR arrays in the planar geometry,
2D anisotropic ($\lambda_x \neq \lambda_y$) SRR arrays in the planar geometry,
and 2D anisotropic ($|\lambda_x| \neq |\lambda_y|$) SRR arrays in the planar-axial
geometry, are shown in Figs. 12(a), 12(b), and 12(c), respectively,
for $\alpha=+1$. These profiles are actually snapshots at some specific 
instant during the DB motion. Here, just like in the 1D case, both the background
and the DB (i.e., the central DB site) oscillate with low and high current
amplitudes, respectively, at the same frequency $\Omega_b=\Omega$.
The stability of the dissipative DBs has been checked by adding small
perturbations of the order of $10^{-2}$ and following their time 
evolution for long time intervals (over $10^{3}T_{b}$ time units). 
In all cases it was found that 
the DBs are not destroyed by the perturbation but, instead, they 
restore their unperturbed shape.
%%%%----------figure12---------
\begin{figure}[!ht]
\includegraphics[angle=0, width=1. \linewidth]{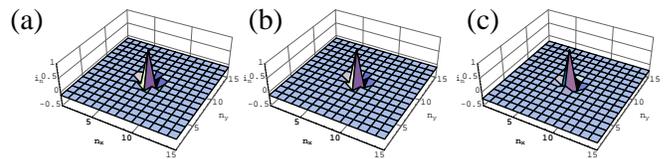}
\caption{(Color online)
Snapshots of two-dimensional dissipative (single-site, bright) 
discrete breathers
(taken at maximum amplitude), for 
$\alpha=+1$, $\Omega_{b}=0.92$, $\varepsilon_0=0.04$, $\gamma=0.01$, 
$\epsilon_\ell=2$ and
(a) $\lambda_{x}=\lambda_{y}=-0.02$ (isotropic split-ring resonator arrays
in the planar geometry);
(b) $\lambda_{x}=-0.02, \lambda_{y}=-0.023$
  (anisotropic split-ring resonator arrays in the planar geometry);
(c) $\lambda_{x}=-0.0124, \lambda_{y}=0.0094$  
  (anisotropic split-ring resonator arrays in the planar-axial geometry).
}
\label{Fig12}
\end{figure}

The magnetic response of the dissipative 2D arrays can be calculated 
as in the 1D case, with the help of Eq. (\ref{13.1}) applied locally
at each array cell $(n,m)$. In Fig. 13 we plot separately the time evolution
of each of the three terms of Eq.(\ref{13.1}), i.e., the instantaneous magnetic 
induction, the applied magnetic field, and the magnetic response,
at the central DB site (Figs. 13(b), 13(d) and 13(f), high amplitude current
oscillation) and the site with $n,m = 3,5$ (Figs. 13(a), 13(c) and 13(e), 
low amplitude current oscillation), for the three DBs shown in Fig. 12. 
The results look the same for the first two cases, corresponding to 
isotropic 2D SRR arrays in the planar geometry (Figs. 13(a) and 13(b)),
and anisotropic 2D SRR arrays in the planar geometry (Figs. 13(c) and 13(d)).
They are also similar with those calculated for the corresponding 1D case. 
The SRRs with low amplitude current oscillation
($(n,m) = (3,5)$) show positive (paramagnetic) response
while the SRR with high amplitude current oscillation (central DB site)
shows extreme diamagnetic (negative) response. 
In the third case, however, for
anisotropic 2D SRR arrays in the planar-axial geometry (Figs. 13(e) and 13(f)),
the response of the SRR with high amplitude current oscillation is still 
diamagnetic but not negative.
This is due to the very weak coupling which presumes relatively large separation
of the SRRs, and thus a very low density SRR array.
%%%%----------figure13---------
\begin{figure}[!t]
\includegraphics[angle=0, width=0.85 \linewidth]{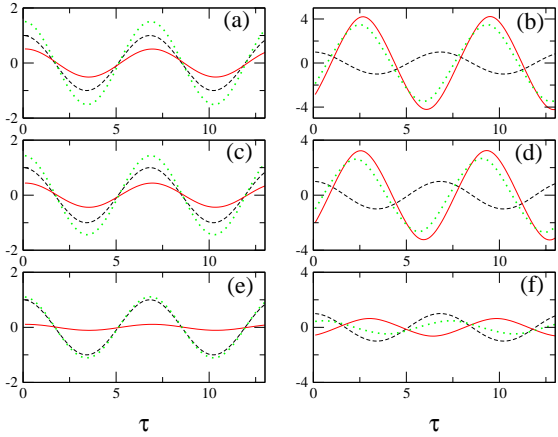}
\caption{(Color online) 
Time evolution of $\ell i_n(\tau)$ (red-solid curve), 
compared with $\cos(\Omega \tau)$ (black-dashed curve), and their sum 
(green-dotted curve), during two breather periods, for a 2D SRR array 
(a) in the planar configuration at $(n,m)=(3,5)$ 
  ($\lambda_x=\lambda_y=-0.02$, $\ell=3$); 
(b) in the planar configuration at the central breather site 
  ($\lambda_x=\lambda_y=-0.02$, $\ell=3$); 
(c) in the planar configuration at $(n,m)=(3,5)$
   ($\lambda_x=-0.02$, $\lambda_y=-0.023$, $\ell=2.6$);
(d) in the planar configuration at the central 
   breather site 
   ($\lambda_x=-0.02$, $\lambda_y=-0.023$, $\ell=2.6$);
(e) in the planar-axial configuration at $(n,m)=(3,5)$
   ($\lambda_x=-0.0124$, $\lambda_y=0.0094$, $\ell=0.43$);
(f)  in the planar-axial configuration at the central breather site
   ($\lambda_x=-0.0124$, $\lambda_y=0.0094$, $\ell=0.43$).
 The other parameters as in Fig. 12.  
}
\label{Fig13}
\end{figure}
Guided from the previous results, we consider the possibility of 
constructing a region in a 2D SRR array with extreme diamagnetic (negative)
response, surrounded  by a paramagnetic background.
For this purpose we may exploit a bright 2D multibreather consisting of a number
of adjacent sites (SRRs) in the center of the array. 
For illustration, such a multibreather occupying a region of nine sites
(including the central one) in an isotropic SRR array in the planar geometry, 
is shown in Fig. 14. We have checked that, indeed, the sites with high 
amplitude current oscillation show negative magnetic response, while the 
rest of them show positive magnetic response. 
Thus, it seems possible to create SRR-based MMs with distinct regions of 
opposite sign magnetic responses by exploiting multibreathers.
%%%%----------figure14---------
\begin{figure}[!h]
\includegraphics[angle=0, width=0.9 \linewidth]{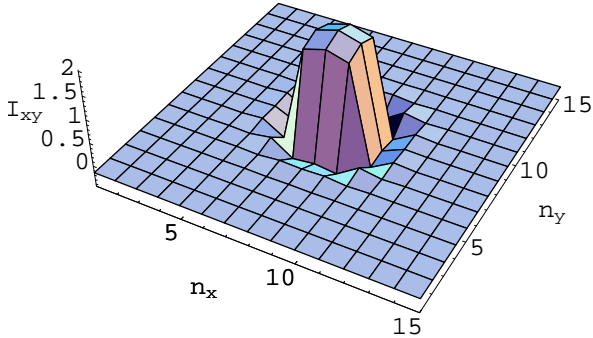}
\caption{(Color online) 
A snapshot of a two-dimensional dissipative multibreather 
(taken at maximum amplitude), constructed for an 
 split-ring resonator array in the planar geometry, for
 $\lambda_{x}=\lambda_{y}=-0.02$, $\omega_{b}=0.92$, $\varepsilon_0=0.04$,
 $\gamma=0.01$, $\epsilon_\ell=2$, $\alpha=+1$.
}
\end{figure}

The numerical results for both 1D and 2D SRR arrays reveal that, 
at least for weak coupling, the amplitude and the time-dependence
of the central DB site and the background are essentially those of the
single damped-driven SRR oscillator. Only the first one or two sites
neighboring to the central DB site exhibit significant differences due
to the coupling. The bistability of the single SRR oscillator,
which is important for the construction of dissipative DBs and also 
for the creation of uniform states with either positive or negative 
magnetic response below resonance, is not restricted to frequencies 
close to resonance. In Fig. 15 we show the time evolution of $i_n$
of two coexisting and stable states far from resonance, along with the
normalized driving magnetic field.  Dissipative DBs can be constructed
by combining these two states into a trivial DB and continuing 
that solution for finite (non-zero) coupling parameters.
Although for the cases shown the current of the high amplitude current
state is rather large, where the saturation of the nonlinear term 
could be expected, Fig. 15 indicates that it is possible to obtain 
negative response below resonance by exciting the SRRs in that state.
%%%%----------figure15---------
\begin{figure}[!t]
\includegraphics[angle=0, width=0.8 \linewidth]{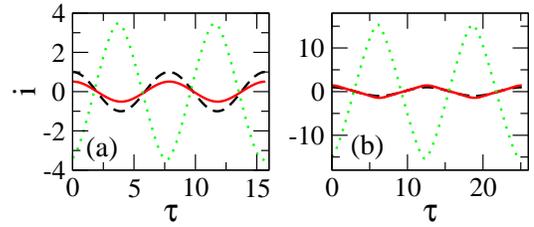}
\caption{(Color online) 
Time evolution of the normalized driving term $\cos(\Omega \tau)$
(black-dashed curve), and the current $i (\tau)$ for the low and high
current amplitude states 
(red-solid and green-dotted curves, respectively) 
for a single damped-driven SRR oscillator with $\gamma=0.01$, $\epsilon_\ell=2$, 
$\alpha=+1$ and
(a) $\Omega =0.8$, $\varepsilon_0=0.2$;
(b) $\Omega =0.5$, $\varepsilon_0=1.2$.
}
\end{figure}

\section{Conclusions
}
We considered a simple model for nonlinear SRR-based MMs in one and two
dimensions, where the nonlinearity arises from a Kerr-type dielectric
which fills the SRR slits. Each SRR, which is modeled as a nonlinear RLC
electrical circuit driven by an alternating voltage source,
is weakly coupled to its nearest neighbors due to magnetic dipole-dipole 
interactions through their mutual inductance $M$ (magnetoinductive coupling).
The sign of the coupling between neighboring SRRs depends on their
relative orientation within the SRR array.   

We have constructed, using standard numerical methods, many different types 
of Hamiltonian and dissipative DBs both in 1D and 2D arrays for different 
nonlinearities (i.e., self-focusing and self-defocusing), and
different geometries (planar and axial in 1D, planar and planar-axial in 2D).
We have also constructed DBs in 2D arrays with moderate anisotropy.
Most of the DBs presented here
are linearly stable under small perturbations. 
Dissipative SRR arrays, driven by an applied magnetic field, 
offer the possibility to study their magnetic response with respect to
that field. The induced current oscillations are proportional to 
the magnetic moments of the SRRs and, thus, to the local magnetization 
(magnetic response) of 
the array. We found that low (high) amplitude current oscillations
are in phase (almost in anti-phase) with the applied field.
Thus, depending on the array and the external field parameters,
the magnetic response at the 
SRRs with high amplitude current oscillation can be negative. 
In this way, DBs can change locally the magnetic response of the array
from paramagnetic to extreme diamagnetic.
By exploiting multibreathers, it seems possible to create SRR-based MMs 
with distinct regions of opposite sign magnetic responses. 

The bistability due to the nonlinearity is not restricted to 
frequencies close to resonance, but it is found to persist down to 
much lower frequencies.  
As a result, dissipative DBs with frequencies and external field amplitudes
in rather wide intervals can be constructed.
Moreover, even far from resonance, 
the coexisting and stable low and high current amplitude states
are in phase and in anti-phase, respectively, with the applied magnetic field.
Thus, it seems possible to get uniform solutions at these low frequencies,
which provide either positive or negative magnetic response below resonance.
This could be achieved by exciting all the SRRs in the low or high amplitude 
state, respectively.

\section*{Acknowledgments}
We acknowledge  support from the grant "Pythagoras II" (KA. 2102/TDY 25)
of the Greek Ministry of Education and the European Union.

\end{document}